\documentclass[12pt,a4paper]{article} 
\pdfoutput=1
\usepackage[whole]{bxcjkjatype} 

\usepackage{amsmath}
\usepackage{amssymb}
\usepackage{bm}
\usepackage{color}
\usepackage{cite} 

\usepackage{hyperref}
\usepackage{graphicx}

\catcode `@=11 \@addtoreset{equation}{section} \catcode `@=12

\begin{document}
\setlength{\oddsidemargin}{0cm}
\setlength{\baselineskip}{7mm}

\begin{titlepage}

	\begin{center}
		{\LARGE
		Constraining Spacetime Dimensions\\ in Quantum Gravity \\
		by Scale Invariance and Electric-Magnetic Duality
		}
	\end{center}
	\vspace{0.2cm}
	\baselineskip 18pt 
	\renewcommand{\thefootnote}{\fnsymbol{footnote}}

	\begin{center}

		Takeshi {\sc Morita}$^{a,b}$\footnote{%
			E-mail address: morita.takeshi(at)shizuoka.ac.jp
		}

		\renewcommand{\thefootnote}{\arabic{footnote}}
		\setcounter{footnote}{0}
		
		\vspace{0.4cm}
		
		{\it
			a. Department of Physics,
			Shizuoka University \\
			836 Ohya, Suruga-ku, Shizuoka 422-8529, Japan 
			\vspace{0.2cm}
			\\
			b. Graduate School of Science and Technology, Shizuoka University\\
			836 Ohya, Suruga-ku, Shizuoka 422-8529, Japan
			\vspace{0.2cm}
		}

	\end{center}
	
	
	\vspace{1.5cm}
	
\begin{abstract}
	We consider a low energy effective theory of $p$-branes in a $D$-dimensional spacetime, and impose two conditions: 1) the theory is scale invariant, and 2) the electric-magnetic dual $(D-p-4)$-branes exist and they obey the same type of interactions to the $p$-branes. (We also assume other natural conditions such as Lorentz invariance but not string theory, supersymmetry, supergravity and so on.) We then ask what $p$ and $D$ are consistent with these conditions. Using simple dimensional analysis, we find that only two solutions are possible: $(p,D)=(2,11)$ and $(p,D)=(2n-1,4n+2)$, ($n=1,2,3,\cdots$). The first solution corresponds to M-theory, and the second solutions at $n=1$ and $n=2$ correspond to self-dual strings in little string theory and D3-branes in type IIB superstring theory, respectively, while the second solutions for $n \ge 3$ are unknown but would be higher spin theories. Thus, quantum gravity (massless spin two theory) satisfying our two conditions would only be superstring theories, and the conditions would be strong enough to characterize superstring theories in quantum gravity.
		\end{abstract}
	
\end{titlepage}


\section{Introduction}
\label{sec-intro}

The construction of the underlying quantum gravity theory in our universe is an important goal in theoretical physics, and the most promising candidate is superstring theories \cite{Green:1987sp, Polchinski:1998rq, Polchinski:1998rr}.
One remarkable property of superstring theories is that spacetime is constrained to ten dimensions\footnote{In this article, we consider theories with perturbatively stable Lorentz invariant vacua. Thus, we do not consider bosonic string theories or non-critical string theories. }. 
Furthermore, there may be a theory of two-dimensional membranes \cite{Bergshoeff:1987cm} in an eleven-dimensional spacetime (M-theory) that unifies the five superstring theories \cite{Witten:1995ex, Schwarz:1995jq, Duff:1999ys}. 
Therefore, the ten and eleven dimensions have a special significance in quantum gravity.
There are several derivations of these dimensions such as the analysis of the maximum supersymmetry \cite{Cremmer:1978km}, supersymmetric extended objects \cite{Bergshoeff:1987cm}, supergravity-like dilaton gravities \cite{Duff:1993ye, Duff:1994an}, and the anomaly cancellation on the world-sheets of superstrings \cite{Green:1987sp, Polchinski:1998rq, Polchinski:1998rr}. 
However, these derivations are mathematically sophisticated, and it is not that obvious why ten and eleven are important.
Moreover, it is not clear whether other candidates for quantum gravity are possible or not.

In this article, we suggest that scale invariance and electric-magnetic duality may provide clues to answer these questions.
We assume that the underlying theory of quantum gravity is defined in a $D$-dimensional spacetime and elementary objects are a $p$-brane and its electric-magnetic dual $(D-p-4)$-brane so that the branes are quantized by the Dirac quantization condition.
By imposing several natural assumptions of quantum gravity, including scale invariance, simple dimensional analysis tells us that only two-brane and five-brane in $D=11$ and a self-dual three-brane in $D=10$ would be consistent with conventional gravity.
These solutions correspond to superstring theories, and it means that scale invariance and electric-magnetic duality may be strong enough to characterize superstring theories.
This derivation of the ten and eleven dimensions is not only simple, but also gives us a novel understanding of superstring theories in quantum gravity.

\section{Derivation of spacetime dimensions}
\label{sec-setup}

We assume that the underlying theory of quantum gravity is defined in a $D$-dimensional spacetime and an elementary object is a $p$-brane, where $p$ is the number of the spatial dimensions of the brane ($D \ge 2$ and $p \ge 0$).
We discuss what $p$ and $D$ are consistent by considering a low-energy effective theory of the branes under certain assumptions of quantum gravity.

\begin{figure}
    \begin{center}
                    \includegraphics[scale=1]{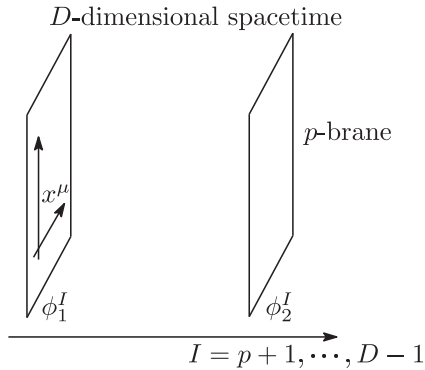}
        \caption{
			Two $p$-branes nearly parallel in a $D$-dimensional spacetime.
        }
        \label{fig-brane}
    \end{center}
\end{figure}

We put two $p$-branes nearly parallel and far enough separated (Fig.~\ref{fig-brane}). 
Then, this system at low energy would be described by the following effective theory,
\begin{align} 
	S_{\text{eff}}=  \int d^{p+1}x \left\{ \sum_{i=1}^2  \left(  \partial \phi^I_i \right)^2 + {\mathcal L}_{\text{int}} \right\},
	\label{action-eff}
\end{align}
where we have taken natural units $c=\hbar=1$.
Here, $x^\mu$ ($\mu=0, \cdots,p $) are the coordinates on the world volume of the $p$-branes and $\phi^I_i(x)$ ($I=p+1,\cdots, D-1$, $i=1,2$) are the target space coordinates of the $i$-th brane in the $D$-dimensional spacetime.
Since we will consider scale invariance later, $\phi^I_i$ are canonically normalized for convenience and their mass dimension is given by
\begin{align} 
	\left[\phi^I_i \right]  = \frac{1}{2}(p-1).
	\label{eq-phi-mass}
\end{align}
Note that we are interested in the low energy physics and the non-relativistic limit has been taken in the effective action \eqref{action-eff} \footnote{The kinetic terms \eqref{action-eff} can be obtained from the conventional world volume action $S=\int$(volume of the $p$-brane) by taking the static gauge and the non-relativistic limit. 
In this paper, we do not consider tensionless branes or massless point particles, which are always relativistic.
This is because the tensionless branes would correspond to massless higher spin theories.
Also, the construction of quantum gravity using massless elementary point particles such as graviton is known to be difficult.
}.
Also, since we will study dimensional analysis, we have omitted the numerical factors of the coefficients and indexes of the derivatives.
Besides, generally other fields such as gauge fields or fermions may exist on the branes, but we omit them and focus on the $\phi^I_i$ fields.

\begin{figure}
    \begin{center}
                    \includegraphics[scale=0.5]{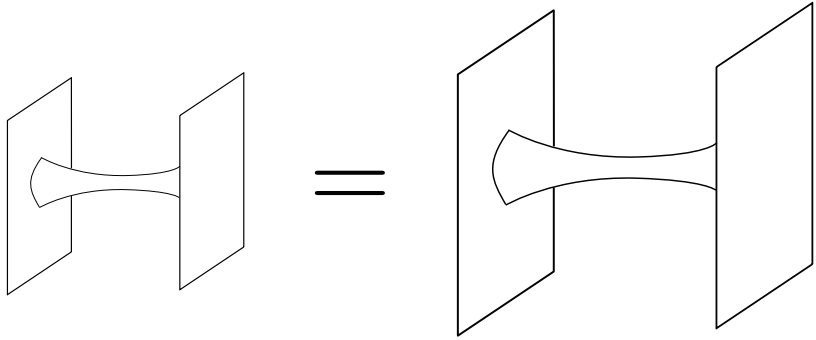}
        \caption{
		Scale invariant interaction between the two $p$-branes.
		We cannot distinguish between ``small" and ``large", if the system is scale invariant.  
		}
        \label{fig-scale}
    \end{center}
\end{figure}

Let us consider possible interaction terms ${\mathcal L}_{\text{int}}$ between the two nearly parallel $p$-branes in the low energy effective theory \eqref{action-eff}.
For simplicity, we define $\phi:=| \phi_1 - \phi_2|$ and focus on the radial mode of the relative motion of the two $p$-branes by considering the situation where other modes and fields excite much more weakly.

Also, we impose the following assumptions on the effective theory,
\begin{description}
	\item[Assumption 1] The system obeys the $D$-dimensional Poincar${\rm \acute{e}}$  symmetry.
	\item[Assumption 2] The effective Lagrangian is non-singular at $\partial^m \phi=0$, ($m=1, 2, \cdots $), and the derivative expansion works (higher derivatives are less relevant).
\end{description}
These are reasonable assumptions, but they are not enough to restrict the theory.
Hence, as a strong constraint, we assume the condition on scale invariance,
\begin{description}
	\item[Assumption 3 (scale invariance)] The system is scale invariant and no dimensionful coupling exists in the Lagrangian \eqref{action-eff}.
\end{description}
Since scale invariance is one of beautiful symmetries in quantum theory (Fig.~\ref{fig-scale}), it may be natural to expect that the underlying theory of our universe respects it.

Now, the leading interaction terms of the effective action \eqref{action-eff} satisfying these three assumptions are given by
\begin{align} 
	{\mathcal L}_{\text{int}}	\sim   
 \frac{\left( \partial_\mu \phi  \partial^\mu \phi \right)^{n}}{\phi^X}  + \cdots.
 \label{action-int}
\end{align}
Here, the power $n$ is restricted to be a non-negative integer by Assumption 2, and the derivative expansion has been considered and  ``$\cdots$" denotes higher order terms. 
The power $X$ is determined by dimensional analysis as
\begin{align} 
	X= 2(n-1)+ \frac{4(n-1)}{p-1}.
\label{X}
\end{align}
($X$ is not fixed when $p=1$, since $[\phi]=0$, and this case is discussed separately in Appendix \ref{app-calculation}.)
If $X$ is an integer, the factor of $1/\phi^X$ in the interaction \eqref{action-int} can be regarded as a conventional long range interaction between two separated parallel $p$-branes in the $D=X+p+3$ dimensional spacetime induced by bulk massless modes \footnote{When $X=0$, the factor of $1/\phi^X$ may be replaced by $\log \phi$. However, it is possible only when $p=1$ and $[\phi]=0$ from Assumptions 2 and 3. Hence, we consider this case in Appendix \ref{app-calculation}. \label{ftnt-X=0}}.
Hence, we impose,
\begin{description}
	\item[Assumption 4] $X$ defined in Eq.~\eqref{X} should be an integer, and the spacetime dimension $D$ is fixed by $D=X+p+3$.
\end{description}
This assumption constrains $n$ and $p$, and we obtain non-trivial relations between $p$ and $D$ \footnote{In the $n=2$ case, we obtain three solutions \cite{Morita:PC}:  $(p,D)=(2,11)$, $(5,11)$ and $(3,10)$. 
They correspond to M2-branes and M5-branes in M-theory and D3-branes in type IIB superstring theory.
Actually, the interaction \eqref{action-int} at $n=2$ naturally arises in maximally supersymmetric conformal field theories on the branes \cite{Maldacena:1997re}.
Thus, if we assume a similar symmetry and restrict $n=2$, these three solutions are our final answer.
Therefore, the condition $n=2$ is powerful enough to extract M-theory and type IIB superstring theory.
A related analysis by using a dilaton gravity theory has been done in Refs.~\cite{Duff:1993ye, Duff:1994an}.
}.
For example, if $n=0$, only $(p,D)=(0,5)$ is allowed. 

However, we still have many possible solutions $(n,p)$ and we impose an additional assumption,
\begin{description}
	\item[Assumption 5 (electric-magnetic duality)] If a $p$-brane exists in the theory, its electric-magnetic dual $(D-p-4)$-brane also exists.
	The interactions between two dual branes are also described by Eq.~\eqref{action-int} with the same power of $n$.
\end{description}
This assumption is also reasonable.
If a $p$-brane exists, we naturally expect that it couples to a $(p+1)$-form field similar to string theories, and, to quantize the charge through the Dirac quantization, its dual $(D-p-4)$-brane that couples to the dual $(D-p-3)$-form field should be involved in the theory \cite{Polchinski:2003bq}. 
(The existence of the dual brane is discussed in the context of the swampland conjecture, too \cite{Agmon:2022thq}.)
Then, if the $p$-branes and the dual branes obey the same dynamics, we expect that the common power of $n$ appears in the interaction \eqref{action-int}, although this point is more speculative \footnote{
	If we relax the condition on $n$ and allow different $n$ for the dual branes, Assumption 5 is still restrictive, but weakened.
	}.

Assumption 5 is expressed by the following equations,
\begin{description}
	\item[Assumption 5'] $(n,p)$ satisfies the relations:
	\begin{align} 
		X&=2(n-1)+ \frac{4(n-1)}{p-1}: ~\text{integer} ,
		\label{eq-npx}
		\\
		\tilde{X}&:=2(n-1)+ \frac{4(n-1)}{\tilde{p}-1}: ~\text{integer}, \quad (\tilde{p}:=D-p-4) ,
		\label{eq-npx'}
		\\
		D&=X+p+3=\tilde{X} +\tilde{p} +3, ~ (D \ge 2~\text{and}~p,~\tilde{p} \ge 0).
		\label{eq-D}
	\end{align}	
\end{description}
Here, $\tilde{p}$ and $\tilde{X}$ are the quantities for the dual brane corresponding to $p$ and $X$ of the original $p$-brane.
We can solve these equations easily as shown in Appendix \ref{app-calculation}, and there are only two solutions\footnote{In our derivation, quantum mechanics is used only in two places. One is that the Planck's constant $\hbar$ is used to relate the energy scale to the length scale in dimensional analysis. The other is in the Dirac quantization. In this sense, our analysis is semi-classical. }:
\begin{align} 
&	\text{\bf{Solution 1:}}\quad
	n=2, \quad  D=11,  \quad (p,\tilde{p})=(2,5), 
	\quad (\text{or equivalently }(p,\tilde{p})=(5,2)  ) ,
	\label{sol-D11}\\
&	\text{\bf{Solution 2:}}\quad
 D=4n+2,  \quad p=\tilde{p}=2n-1, \quad  (n=1,2,3,\cdots).
	\label{sol-SD}
\end{align}
The first solution corresponds to M-theory ($p=2$ and 5 are M2-branes and M5-branes, respectively).
The $n=2$ interaction \eqref{action-int} is also known to be consistent with the supergravity \cite{Maldacena:1997re, Morita:2014ypa}.

The second solution \eqref{sol-SD} is characterized by $p=\tilde{p}$.
In the obtained dimension $D=4n+2$, the self-dual  $(p+2)$-form field strength is possible, where $p+2=2n+1=D/2$ by Eq.~\eqref{sol-SD}, and we presume that the branes are self-dual.
Actually, when $n=2$, we obtain $(p,D)=(3,10)$ corresponding to D3-branes in type IIB superstring theory that are self-dual.
Also, when $n=1$, we obtain $(p,D)=(1,6)$ corresponding to self-dual strings \cite{Gustavsson:2002nz, Kitazawa:2006hq} in little string theory \cite{Seiberg:1997zk, Berkooz:1997cq, Losev:1997hx, Aharony:1999ks}.
The $n=1$ \cite{Gustavsson:2002nz, Kitazawa:2006hq} and 2 \cite{Polchinski:1998rr, Morita:2014ypa} interactions are also consistent with string theories.
(Notice that little string theory is a non-gravitational theory. In our analysis, we have not used genuine gravity, and it is not surprising that the non-gravitational theory is obtained.)
The theories corresponding to the solutions \eqref{sol-SD} for $n \ge 3$ ($D\ge 14$) are unknown\footnote{Although $
(p,D)=(11,26)$ appears at $n=6$, the interaction between two D11-branes in 26 
dimensional bosonic string theory does not agree with the effective Lagrangian 
\eqref{action-int} at $n=6$. So the solution \eqref{sol-SD} at $n=6$ does not 
correspond to bosonic string theory.}.
However, the self-dual field that couples to the self-dual brane  
causes a gravitational anomaly \cite{Alvarez-Gaume:1983ihn}, and it can be 
cancelled only for $D \le 10$ in conventional gravitational theories.
Thus, the solutions for $n \ge 3$ would not correspond to conventional gravitational 
theories. (If the solutions for $n \ge 3$ are not self-dual, they can be 
gravitational theories, although it is unnatural, since the solutions $n=1$ and 2 
would be self-dual.)
Therefore, if the solution \eqref{sol-SD} is self-dual, only $(p,D)=(3,10)$ corresponds to a conventional gravitational theory.

In this way, we have obtained the scale invariant branes and the spacetime dimensions in superstring theories $\{$$(p,D)=(3,10)$, $(2,11)$ and $(5,11)$$\}$ as the solutions satisfying our five assumptions of quantum gravity and corresponding to conventional gravitational theories.
Although the obtained properties of our brane systems are just a part of the properties of the corresponding branes in superstring theories, our systems should be identified with superstring theories, because the knowledge of string theory tells us that it is highly unlikely that there are consistent theories of these branes other than superstring theories.
Therefore, the assumptions proposed in our study are strong enough to characterize superstring theories.

Since we have obtained the scale invariant branes in superstring theories, we can derive other branes and fundamental strings by using string dualities.
It would be instructive to review the derivation of type IIA superstrings from the M2-branes $(p,D)=(2,11)$ \cite{Witten:1995ex, Duff:1987bx}.
It is useful to rewrite the scalars $\phi^I = y^I/l^{3/2} $, where $l$ is a constant and $y^I$ are coordinates of the target spacetime, and both $l$ and $y^I$ have a dimension of length.
Then, the effective action \eqref{action-eff} at $p=2$ becomes
\begin{align} 
	S_{\text{M2}}\sim & \int d^{3}x \left\{ \sum_{i=1}^2 \frac{1}{l^3}  \left( \partial y^I_i \right)^2 
	+ l^9 \left( \frac{1}{l^3} \right)^2 \frac{\left( \partial_\mu  y \partial^\mu y \right)^{2}}{y^6} \right\} 
	+\cdots, 
	\label{action-M2}
\end{align}
where $y:=l^{3/2} \phi$.
In our dimensional analysis, it is possible that an additional dimensionless constant appears as a coefficient of the interaction term, but it is known that no such constant exists in M-theory.
From this expression, we see that $1/l^3$ and $l^9 $ represent the tension of the M2-brane and the $D$-dimensional Newton's constant $G_D$ at $D=11$, respectively. 
(Recall that $[G_D]= 2-D$, which is consistent with $G_{11} \sim l^9$ at $D=11$, and $G_D \times$(tension)${}^2$ is a proper coefficient for the interaction of two $p$-branes.)
Now, we compactify one of the brane direction $x \sim x +R$ where $R$ is the period which breaks the scale invariance.
By introducing constants $l_s$ and $g_s$ through the relations 
$R=g_s l_s$ and $l_s^2= l^3/R$, where $[l_s]=-1$ and $[g_s]=0$, we obtain the effective action after the Kaluza-Klein reduction as, 
\begin{align} 
	S_{\text{M2}}  \sim & \int d^{2}x \left\{ \sum_{i=1}^2 \frac{R}{l^3}  \left( \partial y^I_i \right)^2 
	+ \frac{l^9}{R}  \left( \frac{R}{l^3} \right)^2 \frac{\left( \partial_\mu  y \partial^\mu y \right)^{2}}{y^6} \right\} 
	+\cdots\nonumber \\
	 \sim & \int d^{2}x \left\{ \sum_{i=1}^2 \frac{1}{l_s^2}  \left( \partial y^I_i \right)^2 
	+ g_s^2 l_s^8  \left( \frac{1}{l_s^2} \right)^2 \frac{\left( \partial_\mu  y \partial^\mu y \right)^{2}}{y^6} \right\} 
	+\cdots.
	\label{action-IIA}
\end{align}
This action shows $p=1$ and $X=6$, and it describes a $(p,D)=(1,10)$ system. The tension and the Newton's constant $G_{10}$ are given by $1/l_s^2$ and $g_s^2 l_s^8$, respectively.
This system indeed corresponds to the fundamental strings in type IIA superstring theory, where $l_s$ and $g_s$ represent the string length and the string coupling.
Similarly, we can extract other strings and branes through the string dualities. 
(See, for example, appendix of Ref.~\cite{Morita:2014ypa} for concrete derivations of the brane effective actions.)

\section{Discussions}
\label{sec-discussion}

We have studied the low energy effective theory of two $p$-branes in a $D$-dimensional spacetime, and asked what $p$ and $D$ satisfy the five assumptions proposed for quantum gravity.
We have shown that the assumptions of the scale invariance and the electric-magnetic duality strongly constrain $(p,D)$, and only the two solutions \eqref{sol-D11} and \eqref{sol-SD} are possible.
If the second solution \eqref{sol-SD} is self-dual, just the $D=10$ case can be a gravitational theory (massless spin two theory), and both \eqref{sol-D11} and \eqref{sol-SD} at $D=10$ correspond to superstring theories.
It means that quantum gravities satisfying our assumptions would be superstring theories only. 
As we know, superstring theories exhibit several important properties, and the scale invariance and the electric-magnetic duality might be thought to be only two of them.
However, our study shows that these two may be essential and they are strong enough to extract superstring theories.
Also, M-theory is derived as an exceptional solution \eqref{sol-D11}, and it may emphasize the uniqueness of M-theory.

Assumptions 1, 3 and 5 we have made in this paper are based on the philosophy that the quantum gravity describing our universe is as simple a theory as possible. Assumptions 1 and 3 are made from the point of view that the system will have as much symmetry as possible. Assumption 5 requires that the fundamental objects are subject to the Dirac quantization, otherwise there could be many objects with unrestricted charges in the system, which may complicate the theory. (Note that Assumption 2 and 4 are the conditions that the system is described by the low energy effective theory through the derivative expansions, which is different from the simplicity of the theory.)

The ten and eleven dimensions in superstring theories and supergravities are related to supersymmetries \cite{Green:1987sp, Polchinski:1998rr, Cremmer:1978km, Bergshoeff:1987cm} \footnote{The ten and eleven dimensions are also extracted by considering a $D$-dimensional $p+1$-form potential theory coupled to a dilaton and Einstein gravity \cite{Duff:1993ye, Duff:1994an}. 
This theory is a generalization of supergravity but it is not supersymmetric in general.
This derivation of the spacetime dimensions also does not rely on supersymmetry as does our analysis.
However, our analysis does not require the details of the theory, and can be applied to more general situations and provides the stronger constraints on the spacetime dimensions.
}. 
However, there seems to be no clear answer to the question of why the supersymmetries are essential in our universe. 
On the other hand, our derivation of the ten and eleven dimensions is mainly based on the simplicity of the theory.
Thus, simplicity and supersymmetries may be related, and it would be interesting to explore this relationship further.

We should emphasize that, although the ten and eleven dimensions in superstring theories are obtained in our analysis, they are derived as the dimensions of the scale invariant brane systems rather than those of string theories. 
We have to use the knowledge of string theory and string dualities to reproduce all superstring theories from our scale invariant brane systems. 
Consequently, we cannot obtain bosonic string theory in a 26 dimensional spacetime from our analysis. 
On the other hand, our assumptions of quantum gravity, including the one about the scale invariance, sound natural because of the simplicity. 
Thus, if superstring theory is the underlying theory of our nature, one might claim that superstring theory is chosen but not bosonic string theory, because our nature prefers the natural assumptions.\\

Finally, the theories corresponding to the solution \eqref{sol-SD} with $D \ge 14$ ($n \ge 3 $) are unknown.
One possibility is that these solutions are simply unphysical. 
Another possibility is that they correspond to self-dual branes in some higher spin theories \cite{Vasiliev:1990en, Vasiliev:2003ev, Skvortsov:2018jea, Didenko:2021vui, Bekaert:2022poo}.
If this is the case, since the systems are scale invariant, it may suggest the existence of a correspondence between the higher spin theory on the AdS${}_{2n+1}\times S_{2n+1}$ spacetime and the $2n$ dimensional CFT on the $(2n-1)$-branes similar to the AdS/CFT correspondence \cite{Maldacena:1997re}.
It may be valuable to pursue this possibility.

\paragraph{Acknowledgements}
	The author would like to thank Tetsutaro Higaki, Hikaru Kawai, Toby Wiseman and Tamiaki Yoneya for valuable discussions and comments.
	The author is especially grateful to Shotaro Shiba, Toby Wiseman and Benjamin Withers for discussions on the effective action \eqref{action-int} at $n=2$ \cite{Morita:PC, Morita:2014ypa} that motivated the author to begin this work.
	The work of T.~M. is supported in part by Grant-in-Aid for Scientific Research C (No. 20K03946) from JSPS.

\appendix

\section{Derivation of Eqs.~\eqref{sol-D11} and \eqref{sol-SD} }
\label{app-calculation}

In this appendix, we solve Assumption 5' and derive the solutions \eqref{sol-D11} and \eqref{sol-SD}.
We consider $p=1$ and $p \neq 1$ separately, since $[\phi]=0$ at $p=1$, and $X$ is not determined by Eq.~\eqref{X}. 

\subsection{The $p\neq 1$ case}

By substituting Eq.~\eqref{eq-npx} into the relation $D=X+p+3$, we obtain
\begin{align} 
&	-(p-1)D+2n(p+1)+p^2-5=0, 
\label{eq-npXD} \\
& -(\tilde{p}-1)D+2n(\tilde{p}+1)+\tilde{p}^2-5=0.
\label{eq-npXD-dual}
\end{align}
Here, Eq.~\eqref{eq-npXD-dual} is obtained by replacing $p \to \tilde{p}$ in Eq.~\eqref{eq-npXD}.
By subtracting these two equations, we obtain
\begin{align} 
	0=	(\tilde{p}-p)\left(D-2n-p-\tilde{p}\right)=(D-2p-4)(4-2n),
\end{align}
and the solution is given by $D=2p+4$ or $n=2$. 

For the first solution $D=2p+4$, by substituting it into Eq.~\eqref{eq-npXD}, we obtain
\begin{align} 
	(2n-p-1)(p+1)=0.
\end{align}
Since $p \ge 0$, the solution is given by $p=2n-1$.
Then, $\tilde{p}=D-p-4=p$, and the solution \eqref{sol-SD} except $p = 1$ ($n=1$) is obtained.

For the solution $n=2$ \cite{Morita:PC}, by substituting it into Eq.~\eqref{eq-npXD}, we obtain
\begin{align} 
	D=p+5 + \frac{4}{p-1}.
\end{align}
Since $D$ is an integer greater than 2, we obtain $p=2,3$ and $5$.
For $p=2$ and 5, we obtain $D=11$ and they are the solution \eqref{sol-D11}.
For $p=3$, $D=10$ ($ \tilde{p} =3$) is obtained, and it is a solution of Eq.~\eqref{sol-SD}.

\subsection{The $p = 1$ case}
When $p = 1$, the mass dimension of $\phi$ becomes 0 from \eqref{eq-phi-mass}.
Then, the mass dimension of the interaction \eqref{action-int} is $2n$ for any $X$, and we obtain $n=1$ to make the action dimensionless.
However, dimensional analysis cannot determine the value of $X$ and $D$, and we fix them by using the electric-magnetic duality (Assumption 5).
When $n=1$, only $p=1$ is possible to make the action dimensionless ($X=0$ is another possibility but it also requires $p=1$. 
See footnote \ref{ftnt-X=0}). 
Therefore, the electric-magnetic dual of the $p=1$ brane should be an 1-brane ($\tilde{p}=1$). 
It implies $D=6$ ($X=2$) and we obtain the solution \eqref{sol-SD} for $n=1$.

{\normalsize 
\bibliographystyle{utphys}
\bibliography{string}

}

\end{document}